\documentclass[preprint]{revtex4-1}
\usepackage{graphicx}
\usepackage{amsmath}
\usepackage{xcolor}

\begin{document}


\title{Inelastic scattering of electrons in water from first-principles: cross sections and inelastic mean free path for use in Monte Carlo track-structure simulations of biological damage}

\author{Natalia E. Koval}
 \email{natalia.koval.lipina@gmail.com}
 \affiliation{CIC Nanogune BRTA, Tolosa Hiribidea 76, 20018 Donostia-San Sebasti{\'a}n, Spain}
 
  \author{Peter Koval}
  \email{koval.peter@gmail.com}
 \affiliation{Simune Atomistics S.L., Tolosa Hiribidea 76, 20018 Donostia-San Sebasti{\'a}n, Spain}
 
 \author{Fabiana Da Pieve}
 \affiliation{Royal Belgian Institute for Space Aeronomy BIRA-IASB, 1180 Brussels, Belgium}
 
 \author{Jorge Kohanoff}
 \affiliation{Atomistic Simulation Centre, Queen’s University Belfast, Belfast BT71NN,
Northern Ireland, United Kingdom}
\affiliation{Instituto de Fusion Nuclear "Guillermo Velarde", Universidad Politecnica de Madrid, 28006 Madrid, Spain}
 
 \author{Emilio Artacho}
 \affiliation{CIC Nanogune BRTA and Donostia International Physics Center DIPC, Tolosa Hiribidea 76, 20018 Donostia-San Sebasti{\'a}n, Spain}
\affiliation{Ikerbasque, Basque Foundation for Science, 48011 Bilbao, Spain}
\affiliation{Theory of Condensed Matter, Cavendish Laboratory, University of Cambridge, J. J. Thomson Ave, Cambridge CB3 0HE, United Kingdom}

 \author{Dimitris Emfietzoglou}
 \affiliation{Medical Physics Laboratory, University of Ioannina Medical School, GR-45110 Ioannina, Greece}





\date{\today}

\begin{abstract}
Modelling the inelastic scattering of electrons in water is fundamental, 
given their crucial role in biological damage. In Monte Carlo 
track-structure codes used to assess biological damage, the
energy loss function, from which cross sections are extracted, is derived 
from different semi-empirical optical models.
Only recently, first \emph{ab-initio} results for the energy loss function and cross-sections in water became available. For benchmarking purpose, in this work, we present \emph{ab-initio} linear-response time-dependent density functional theory calculations of the energy loss function of liquid water. We calculated the inelastic scattering cross sections, inelastic mean free paths, and  electronic stopping powers and compared our results with recent calculations and experimental data showing a good agreement. In addition, we provide an in-depth analysis of the contributions of different molecular orbitals, species, and orbital angular momenta to the total energy loss function. Moreover, we present single-differential cross sections computed for each molecular orbital channel, which should prove useful for Monte-Carlo track-structure simulations.
\end{abstract}


\maketitle 


\section{\label{sec:intro} Introduction}

The scattering of electrons in biological matter plays a crucial role in
a variety of fields related to radiation-induced damage, such as 
ion-beam therapy and risk assessment in space radiation studies.
In theoretical and often experimental studies of biological damage, 
liquid water is considered a model system.
The initial
response of biological material to radiation is determined, to a large
extent, by the oscillator-strength distribution of its valence
electrons, which leads primarily to the generation of
electrons with energies of less than 100 eV~\cite{Plante_2009,pimblott}.
This results from the 
shape of the differential cross section of molecules, 
which peaks at $20-30$ eV and
decreases to low values above 100 eV~\cite{michaud}.  Evidence has 
accumulated throughout the years that very low-energy ($<$20 eV) 
electrons play a relevant role in 
bio-damage~\cite{bouda,aliza,khor,surdu,denifl}. They
constitute the so-called "track ends" and are reported to have an
increased biological effectiveness~\cite{nik1991,niklind,rabus2011}.
Experiments have indicated that electrons (or photons) with energies as
low as few electron-volts can still induce double-strand breaks (DSBs), possibly  
through a resonance mechanism~\cite{huels,prise,Kohanoff_2017}.

Nowadays, several Monte Carlo track-structure (MC-TS) codes exist~\cite{nikjoo1052,Nikjoo_2016,ding2012,konst}, able to describe
the transport of electrons via an event-by-event simulation until low 
energy ($\sim$10 eV), like {\sc norec}~\cite{seme}, {\sc kurbuc}~\cite{liam2012}, 
{\sc partrac}~\cite{pare}, {\sc ritracks}~\cite{Plante_2008}, and the open
source Geant4-DNA~\cite{inc2018,bern2015}.
The track structures in water are then overlaid onto DNA models ranging 
in complexity from simple cylindrical models of the DNA to a full 
atomistic description of human chromosomal DNA~\cite{Garty_2010,doi:10.1063/1.4802962,10.1093/rpd/ncv143,Margis_2020,devera2021}. MC-TS codes either rely on pre-parameterized sets of 
cross sections~\cite{Plante_2009} or use optical data models for the 
dielectric function of water based on the first Born 
approximation~\cite{ritchie,ritchie2,penn,ding53,emfi2005bis}. In such 
models, the energy loss function (ELF) in the dielectric formalism, from
which other quantities like cross sections and inelastic mean free path (IMFP) are calculated, is 
determined at negligible momentum transfer $\textit{q}=0$ from experimental data (often optical 
data~\cite{wata,haya2000}). For finite momentum transfers $\textit{q}\neq$0, the ELF is 
appropriately extended to the 
whole Bethe surface via dispersion models based on the electron gas
theory~\cite{nikjoo1052} within the context of the random-phase 
approximation (RPA). 
The ELF is commonly described via a superposition of either normal or derivative 
Drude functions. The Drude model parameters associated with the 
height, width, and position of the peaks, respectively, are used as 
adjustable parameters determined by the fit to experimental data and are 
generally constrained by sum rules. Other models exist as well, based on 
Mermin functions~\cite{abril},
which partially incorporate effects beyond RPA.

By assuming that each electron interacts with the average field 
generated by all other electrons, the RPA accounts only for electrostatic 
screening. The
exchange-correlation (XC) effects (due to the 
instantaneous Coulomb repulsion and the Pauli exclusion principle) are 
neglected in RPA.
The Born 
approximation neglects, among other things, exchange effects between the
incident and struck electrons. For high energies, such effects are only 
important for hard collisions, characterized by a large energy transfer.
At low energies, however, the incident and the target electrons
have similar energies and thus, one expects that exchange
effects will become relevant
essentially for all collisions. Moreover, in the Born approximation, a 
first-order perturbation theory is used to describe the interaction 
between the projectile and the target, which is in principle not valid 
for low energies~\cite{RevModPhys.43.297,https://doi.org/10.1002/sia.5878}. The use of
approximations for the scattering parameters leads to
differences in simulations of track structure~\cite{Nikjoo1994}.
The discrepancies in the inelastic scattering obtained with different 
extension algorithms to extrapolate optical data to finite momentum 
transfer can reach about a factor of two in the range $50-200$ eV (and even 
larger at still lower energies)~\cite{Emf2012}. Recent studies have 
reported a potentially relevant effect of the different dielectric 
function implementations on ionization clustering~\cite{villa} and DNA 
damage induction~\cite{lampe2018}.

There is a high degree 
of uncertainty in the low-energy range in MC-TS codes as the cross sections become sensitive
to the details of the electronic structure of the 
target~\cite{Champion_2003,ding122,emfi188,garcia2017_49}. Even though
atomic structures are implemented in some TS 
codes~\cite{fried711,taleei,bug2017}, the absence of electronic effects and
often occurring lack of reference cross sections for benchmarking make it 
difficult to extend the applicability of such codes to the targets other 
than homogeneous liquid water. In fact, codes like Geant4-DNA use the 
cross sections for water independently of the actual medium, only 
re-scaling the density.

Several recent works have been performed to ameliorate the description 
of the dielectric function and the IMFP of water. One 
way is to include the XC effects beyond RPA on the
basis of the electron gas 
model~\cite{https://doi.org/10.1002/sia.5878,10.1667/RR13362.1}. Other 
works additionally tried improving the effects beyond the first Born 
approximation~\cite{emfi122,devera2021}, improving previous dispersion 
algorithms~\cite{Truong_2018}, developing new TS 
codes~\cite{quinto,verk}, and clarifying differences in inelastic 
scattering between different condensed phases~\cite{signorell}. Others
focus on extending the set of cross sections for electron 
scattering in targets other than water via pre-parameterized 
models~\cite{zein} or multi-channel and R-matrix 
approaches~\cite{felipe}.

Overall, the accuracy of the semi-empirical results for water at energies below 100 eV 
remains questionable. \emph{Ab-initio} calculations can provide insightful 
results for the dielectric function, the electron energy loss function,
and the inelastic mean 
free path in the whole energy range. First-principles methods do not rely on any free parameters and thus have predictive 
power and can be extended to a variety of targets. Nowadays, 
time-dependent density functional theory (TDDFT)~\cite{runge,tddft-book} is the 
method of choice for the study of excited states, since it allows for the affordable extraction of 
physical information without \emph{a-priori} assumptions on the system 
and on the knowledge of associated cross sections. TDDFT can be 
formulated either in the perturbative regime~\cite{ulrich} or via 
an explicit solution of the time-dependent Kohn-Sham (KS)
equations~\cite{yab1,yab2} by propagating the KS orbitals in real time. 
Nevertheless, \emph{ab-initio} studies on 
the ELF and IMFP in water for low-energy electrons 
are limited so far. A previous real-time TDDFT study by 
Tavernelli~\cite{PhysRevB.73.094204} presented a dielectric constant
of liquid water (optical limit only) with two prominent peaks as opposed to only 
one main peak in the experimental data~\cite{Emf2012}.
A more recent work by Taioli et 
al.~\cite{Taioli2021} presented linear-response TDDFT (LR-TDDFT) calculations of the
ELF for liquid water. The sample of 32 molecules was obtained from a larger sample generated via classical molecular dynamics simulations and then optimized via DFT. The XC effects were considered
in the adiabatic generalized gradient approximation (AGGA). The ELF obtained in Taioli et 
al.~\cite{Taioli2021} has shown a good agreement with experiments. 
However, the orbital analysis of the ELF has not been performed in the \emph{ab-initio} framework.

In this work, we performed a detailed 
first-principles investigation of the electron scattering in liquid water both
in the optical limit and for finite momentum transfer.
Using an efficient iterative method based on LR-TDDFT and a
linear combination of atomic orbitals (LCAO) incorporated in the {\sc mbpt-lcao} 
code~\cite{KOVAL2015216,mbpt-lcao} we calculated the ELF of liquid water 
in a range of finite values of the momentum transfer.
The electron-electron 
interactions were considered at the RPA level in the linear response, 
unlike in the work by Taioli et al.~\cite{Taioli2021}, who used the 
AGGA.
RPA yields the correct asymptotic behavior for the long-range interactions absent in AGGA~\cite{Ren2012}.
The inelastic 
scattering cross sections, the IMFP, and the electronic stopping 
power of electrons in water were then calculated from the ELF using analytical expressions \cite{10.1667/RR13362.1, emfi2005bis}. Furthermore, we performed a detailed analysis of the
contributions of molecular orbitals, chemical species, and their 
pairs, as well as orbital angular momenta to the ELF. Additionally, we computed the cross sections for different 
molecular orbital channels which can benchmark semi-empirical calculations. Apart from the
results presented in this article, we provide the data at \url{https://doi.org/10.5061/dryad.d51c5b057} for the peruse in MC-TS simulations.

\section{\label{sec:method}Methodology}

\subsection{\label{sec:LR}Linear-response time-dependent density functional theory calculations with {\sc mbpt-lcao}}

The ELF is the fundamental quantity that defines the scattering 
properties of a material. It is defined as the imaginary part of the 
inverse macroscopic dielectric function $\epsilon_\mathrm{M}$ \cite{EMFIETZOGLOU2003373}:
\begin{equation}
   \mathrm{ELF}(E, \textbf{q}) = \mathrm{Im}\displaystyle\left[-1/\epsilon_\mathrm{M}(E, \textbf{q})\right],
\end{equation}
that relates the external 
perturbation (potential) $V^\mathrm{ext}$ and the total potential $V^\mathrm{tot}$ acting in a system: $V^\mathrm{ext}(E, \textbf{q}) = \epsilon_\mathrm{M}(E, \textbf{q}) V^\mathrm{tot} (E, \textbf{q})$.

In LR-TDDFT (see Ref.~\cite{Botti_2007} for a broad overview), the main quantity that gives all the information about the
response of a solid to an external perturbation $V^\mathrm{ext}$ is the microscopic 
dielectric function $\epsilon$.
The macroscopic dielectric function can be obtained from the microscopic one using the so-called macroscopic averaging, i.e., by averaging the 
microscopic quantities over all the unit cells, since macroscopic 
quantities slowly vary over the unit cell while microscopic ones vary rapidly \cite{PhysRev.129.62}:
\begin{equation}\label{eps}
    \epsilon_\mathrm{M}(E, \textbf{q})=
    1/\epsilon^{-1}_{\mathbf{G}=0,\mathbf{G'}=0}(E, \textbf{q}) \neq \epsilon_{\mathbf{G}=0,\mathbf{G'}=0}(E, \textbf{q}).
\end{equation}
Here $\mathbf{G,G'}$ are lattice vectors in the reciprocal space, which
is more convenient to use when dealing with periodic systems.
The differences between microscopic and averaged 
(macroscopic) fields are called the crystal local fields, or local field effects (LFE).

The inverse microscopic dielectric function is related to the 
interacting linear-response function $\chi_\mathbf{{GG}'}(E, \textbf{q}) = \delta n_{\mathbf{G}}(E, \textbf{q})/\delta V^{\mathrm{ext}}_{\mathbf{G'}}(E, \textbf{q})$, an operator producing 
the induced density $\delta n$ in response
to a change of an external potential $\delta V^{\mathrm{ext}}$:
\begin{equation}\label{eps-1}
    \epsilon^{-1}_\mathbf{GG'}(E, \textbf{q}) = \delta_\mathbf{GG'} + \upsilon_\mathbf{GG''}(\textbf{q})  \chi_\mathbf{{G''G}'}(E, \textbf{q}),
\end{equation}
where $\upsilon_\mathbf{GG'}(\textbf{q}) = 4\pi \delta_\mathbf{GG'} / |\textbf{q}+\mathbf{G}|^2$ is the Coulomb interaction matrix element between plane-waves and $\delta_\mathbf{GG'}$ is the Kronecker delta symbol. We adopt the repeated index sum convention in this section.
Thus, the ELF in terms of the interacting response function reads: 
\begin{equation}\label{elf-mbpt}
\text{ELF}(E,\textbf{q}) = -\frac{4\pi}{\textit{q}^2}\text{Im}\chi_{\mathbf{G}=0,\mathbf{G'}=0}(E,\textbf{q}).
\end{equation}
In the KS formalism, the external potential is related to an effective potential 
$V^{\mathrm{eff}}(\textbf{r}, t) \equiv V^{\mathrm{ext}}(\textbf{r}, t) + V^{\mathrm{Hxc}}(\textbf{r}, t)$, where Hxc stands for Hartree+XC potential -- the relation generating a Dyson-type
equation for the interacting $\chi_{\mathbf{GG}'}(E,\textbf{q})$ and non-interacting $\chi^{0}_{\mathbf{GG}'}(E,\textbf{q})$ response functions \cite{PhysRevLett.76.1212}: 
\begin{equation}\label{chi}
 \displaystyle{   \chi_\mathbf{{GG}'}(E, \textbf{q}) =  \chi^0_\mathbf{{GG}'}(E, \textbf{q}) +} \displaystyle{ \chi^0_\mathbf{{GG}''}(E, \textbf{q}) K^\mathbf{G''G'''}(\textbf{q}) \chi_\mathbf{G'''G'}(E, \textbf{q})}.
\end{equation}
The crucial ingredient here is the interaction kernel $K^\mathbf{GG'}(\textbf{q})$, given by:

\begin{equation}\label{kernel1}
  K^\mathbf{GG'}(\textbf{q}) = \upsilon_\mathbf{GG'}(\textbf{q}) + \mathsf{f}_{\mathrm{xc};\textbf{GG'}}(\textbf{q}).
\end{equation}
The XC kernel $\mathsf{f}_\mathrm{xc}$ is the functional derivative of
the time-dependent XC-potential with respect to the time-dependent particle density and 
$\upsilon_\mathbf{GG'}$ is the functional derivative of the Hartree 
potential with respect to the density. In this work, we calculated the LR-TDDFT 
interaction kernel $K^\mathbf{GG'}(\textbf{q})$ in the RPA
approximation $K^\mathbf{GG'}(\textbf{q}) = \upsilon_\mathbf{GG'}$. 
The XC effects are only taken into account in the ground state calculations.
The RPA response function,
only accounting for the Hartree component of the induced potentials, 
generally provides a good 
description of long-range screening \cite{Olsen2019}. 

The non-interacting response function appearing in Eq.~\eqref{chi} can 
be calculated as follows~\cite{PhysRev.126.413,PhysRev.129.62}:
\begin{equation}\label{chi0}
\chi^0_\mathbf{{GG}'}(E, \textbf{q}) = \dfrac{1}{N_\mathrm{k}} \sum_\mathrm{n,m,\mathbf{k}} \dfrac{(\mathit{f}_\mathrm{n,\mathbf{k}}-\mathit{f}_\mathrm{m,\mathbf{k}+\textbf{q}})U^\mathbf{G}_{\mathrm{nm},\mathbf{k}}(\textbf{q})\overline{U}^\mathbf{{G}'}_{\mathrm{nm},\mathbf{k}}(\textbf{q})}{E-(\mathcal{E}_{\mathrm{m},\mathbf{k}+\textbf{q}}-\mathcal{E}_\mathrm{n,\mathbf{k}})+\mathrm{i}\eta}.
\end{equation}
Here, $\mathit{f}_\mathrm{n,\mathbf{k}}$ and $\mathcal{E}_\mathrm{n,\mathbf{k}}$ are the occupation numbers and the energies of the corresponding KS eigenstates, $\eta$ is a broadening constant, and $N_\mathrm{k}$ is the number of k-points in the chosen Brillouin zone (BZ) sampling. $U^\mathbf{G}_\mathrm{nm}(\mathbf{k},\textbf{q})$ are the matrix elements of plane-waves in the basis of KS eigenstates $\Psi_{\mathrm{n,\mathbf{k}}}(\mathbf r)$:
\begin{equation}\label{Unm}
U^\mathbf{G}_{\mathrm{nm},\mathbf{k}}(\textbf{q}) = \dfrac{1}{\sqrt{V_{\mathrm{uc}}}} \int \Psi^*_{\mathrm{n,\mathbf{k}}}(\mathbf r) \mathrm e^{\mathrm{-i}(\mathbf{G}+\textbf{q})\mathbf{r}} \Psi_{\mathrm{m,\mathbf{k}+\textbf{q}}}(\mathbf r) d^3r,
\end{equation}
where $V_{\mathrm{uc}}$ is the unit cell volume.

The {\sc mbpt-lcao} code uses an efficient iterative Krylov-subspace method to calculate the ELF (Eq.~\eqref{elf-mbpt}). More details about the LR-TDDFT implementation in the {\sc mbpt-lcao} 
code can be found in Ref.~\cite{KOVAL2015216}.


\subsection{\label{sec:partition}Partition of the electron energy loss function}

The non-interacting response function $\chi^{0}_{\mathbf{GG}'}(E,\textbf{q})$
has an explicit expression in terms of KS orbitals 
$\Psi_{\mathrm{n},\textbf{k}}(\textbf{r})$, their eigenenergies 
$\mathcal{E}_{\mathrm{n},\textbf{k}}$ and 
occupations $\mathit{f}_{\mathrm{n},\textbf{k}}$ (Eqs.~\eqref{chi0} and \eqref{Unm}).
To analyze the contribution of different orbitals,
as well as different species to the total ELF, we will express the ELF via the 
non-interacting response function. For this, it is convenient to rewrite Eq.~\eqref{chi} in the following form:
\begin{equation}\label{new_chi}
\chi=\chi^0 \left[\delta - K \chi^0 \right]^{-1}.
\end{equation}
The operator $\left[\delta - K \chi^0 \right]^{-1}$ converts the external potential 
$V^{\mathrm{ext}}_{\mathbf{G}+\textbf{q}}(E)$ to an effective potential $V^{\mathrm{eff}}_{\mathbf{G}+\textbf{q}}(E)$.
In our case, $V^{\mathrm{ext}}_{\mathbf{G}+\textbf{q}}(E)\equiv \delta_{\mathbf{G},\mathbf{0}}$ and the effective potential
$V^{\mathrm{eff}}_{\mathbf{G}+\textbf{q}}(E)$ is 
computed by solving 
\begin{equation}
\left[\delta_{GG'} - K_{\mathbf{GG}''} \chi^0_{\mathbf{G}''\mathbf{G}'}
(E,\textbf{q})\right]
V^{\mathrm{eff}}_{\mathbf{G}'+\textbf{q}}(E) =
V^{\mathrm{ext}}_{\mathbf{G}+\textbf{q}}(E).
\end{equation}

In what is described in Sec.~\ref{sec:LR}, the computation of the 
ELF is performed by applying the non-interacting response to the effective 
potential:
\begin{equation}
\text{ELF}(E,\textbf{q}) = -\frac{4\pi}{\textit{q}^2}\text{Im}\chi^0_{\mathbf{G=0,G}'}(E,\textbf{q})
V^{\mathrm{eff}}_{\mathbf{G}'+\textbf{q}}(E).
\end{equation}
Summing over the reciprocal lattice vectors $\textbf{G}'$, we get:
\begin{equation}
\text{ELF}(E,\textbf{q}) = -\frac{4\pi}{\textit{q}^2}\text{Im}
\sum_{\mathrm{n,m},\textbf{k}}
U_{{\mathrm{nm}},{\textbf{k}}}(\textbf{q})
D_{{\mathrm{nm}},{\textbf{k}}}(E,\textbf{q}),
\label{ELF-KS}
\end{equation}
where 
\begin{equation}
U_{\mathrm{nm},{\textbf{k}}}(\textbf{q}) = 
\int \Psi^*_{\mathrm{n},\textbf{k}}(\textbf{r})
\mathrm{e}^{-\mathrm{i}\textbf{q}\textbf{r}}
\Psi_{\mathrm{m},\textbf{k}+\textbf{q}}(\textbf{r}) \, d^3r,
\end{equation}
and 
\begin{equation}
D_{{\mathrm{nm}},{\textbf{k}}}(E,\textbf{q}) = 
\sum_{\mathbf{G}'} (\mathit{f}_{\mathrm{n},\textbf{k}} - \mathit{f}_{\mathrm{m},\textbf{k}+\textbf{q}})
\times \frac{
\displaystyle
\int 
\Psi^*_{\mathrm{m},\textbf{k}+\textbf{q}}(\textbf{r})
\mathrm{e}^{\mathrm{i}(\mathbf{G}' + \textbf{q})\textbf{r}}
\Psi_{\mathrm{n},\textbf{k}}(\textbf{r}) \, d^3r
}{
E - (\mathcal{E}_{\mathrm{m},\textbf{k}+\textbf{q}}-\mathcal{E}_{\mathrm{n},\textbf{k}}) + \mathrm{i}\eta
}V^{\mathrm{eff}}_{\mathbf{G}'+\textbf{q}}(E).
\end{equation}

Since Eq.~(\ref{ELF-KS}) is linear in all the indices,
we can split the ELF into different contributions. 
For example, the contributions of the electron-hole pairs 
can be defined as
\begin{equation}
\text{ELF}_{\mathrm{nm},\textbf{k}}(E,\textbf{q}) = -\frac{4\pi}{\textit{q}^2}\text{Im}
\left[U_{\mathrm{nm},{\textbf{k}}}(\textbf{q})
D_{\mathrm{nm},{\textbf{k}}}(E,\textbf{q}) \right],
\label{ELF-EH}
\end{equation}

In practice, when the (atomistic) system is 
large and consists of many almost equal parts (water model), it is 
desirable to estimate the contributions of different (crystalline) 
orbitals defined by their energies $\mathcal{E}_{\mathrm{n},\textbf{k}}$ to the total $\mathrm{ELF}(E,\textbf{q})$. 
This is achieved by defining an occupied-energy differential ELF (DELF):

\begin{equation}
\text{DELF}^{\mathrm{occ}}(E,\textbf{q},\mathcal{E}) =
    \sum_{\mathrm{n}<\mathrm{m}} \delta(\mathcal{E} - \mathcal{E}_{\mathrm{n},\textbf{k}}) \text{ELF}_{\mathrm{nm},\textbf{k}}(E,\textbf{q})
    + \sum_{\mathrm{m}<\mathrm{n}} \delta(\mathcal{E} - \mathcal{E}_{\mathrm{m},\textbf{k+q}}) \text{ELF}_{\mathrm{nm},\textbf{k}}(E,\textbf{q}).
\end{equation}

The contributions of different angular momenta, species, or 
combination of these can be ``tracked'' using an expansion of the KS orbitals in terms of the atomic orbitals
\begin{equation}
\Psi_{\mathrm{n},\textbf{k}}(\textbf{r}) = \sum_a C^a_{\mathrm{n},\textbf{k}} \phi^{a}(\textbf{r}, \textbf{k}),
\end{equation}
where $C^a_{\mathrm{n},\textbf{k}}$ are the LCAO coefficients and $\phi^{a}(\textbf{r}, \textbf{k})$ are the Bloch-symmetric atomic orbitals.
The atomic orbital index $a$ is connected to particular atoms, angular momenta, etc.

\subsection{\label{sec:CS} Inelastic scattering cross sections and stopping power from the energy loss function}

As has been mentioned above, all the relevant quantities in the inelastic electron scattering can be calculated 
from the ELF~\cite{emfi2005bis,10.1667/RR13362.1}. 
According to the non-relativistic plane-wave Born approximation, the double-differential inelastic scattering cross section
is defined as follows:

\begin{equation}\label{ddcs}
\frac{d^2\Sigma(E,{\textit{q}};T)}{dEd{\textit{q}}} = \frac{1}{\pi a_0 T \textit{q}}
\text{ELF}(E,\textit{q}),
\end{equation}
where $T$ is the incident electron kinetic energy, $a_0$ is the Bohr radius.
Note that here we used
the magnitude $\textit{q}$ of the momentum transfer vector $\textbf{q}$ since it is expected to be isotropic in liquid water (and our 
tests confirm this for our sample, see figure \ref{qdir} in Appendix A).

The single-differential cross section (SDCS) can be obtained from \eqref{ddcs} by integrating over momentum transfer $\textit{q}$:
\begin{equation}\label{sdcs}
\frac{d\Sigma(E;T)}{dE}=\frac{1}{\pi a_0T}\int\displaylimits^{\textit{q}_\mathrm{max}(E;T)}_{\textit{q}_\mathrm{min}(E;T)}
\frac{\text{ELF}(E,\textit{q})}{{\textit{q}}}d{\textit{q}},
\end{equation}
where the limits
$\textit{q}_\mathrm{min/max}(E;T)=\sqrt{2m}(\sqrt{T} \mp \sqrt{T-E})$ 
come from momentum conservation ($m$ is the electron rest mass).

The total inelastic cross section $\Sigma$ (also called inverse IMFP) and the
electronic (or collisional) stopping power $S_\mathrm{e}$ are defined as:

\begin{equation}\label{tot_cs}
\Sigma(T)=\int\displaylimits^{E_\mathrm{max}(T)}_{E_\mathrm{min}}\frac{d\Sigma(E;T)}{dE}dE,
\end{equation}
\begin{equation}\label{stopp}
S_e(T)=-\frac{dT}{dx}=\int\displaylimits^{E_\mathrm{max}(T)}_{E_\mathrm{min}}E \frac{d\Sigma(E;T)}{dE}dE,
\end{equation}
with the maximum energy loss by an electron with energy $T$ being 
$E_\mathrm{max} = \mathrm{min}[(T + E_\mathrm{gap})/2, 
T-E_\mathrm{F}]$ for insulators, where $E_\mathrm{gap}$ is the energy
gap of the target, and $E_\mathrm{F}$ is the Fermi energy; and the minimum defined as $E_\mathrm{min} = E_\mathrm{gap}$ \cite{10.1667/RR13362.1, garcia2017_49}. Notice that this 
assumes that electronic excitation can only occur for energies larger than the 
gap, which acts as an effective threshold. However, recent studies have shown 
that this is possible also for energies below the gap \cite{Artacho_2007,PhysRevLett.116.043201}. Here, we go beyond the threshold, i.e., use the limit $E_\mathrm{min}=0$ in Eqs.~\eqref{tot_cs} and \eqref{stopp}.

\section{Numerical details}

\subsection{\label{sec:siesta} {\sc siesta} single-point and {\sc mbpt-lcao} calculations}

The ground state KS orbitals of the water samples needed as a starting 
point for the LR-TDDFT calculations, were obtained using the static 
DFT as implemented in the {\sc siesta} code~\cite{siesta} using periodic 
boundary conditions.
A $2\times2\times2$ 
Monkhorst-Pack~\cite{PhysRevB.13.5188} $k$-point mesh was used in the 
{\sc siesta} calculations. The XC functional in the 
local-density approximation (LDA) in the Ceperley-Alder 
form~\cite{PhysRevLett.45.566} was used. Norm-conserving 
Troullier-Martins~\cite{PhysRevB.43.1993} 
pseudopotentials were used to replace the core electrons. Basis sets of different
sizes, i.e., single-, double-, and triple-$\zeta$ polarized (SZP, DZP 
and TZP, respectively) with an energy shift of 20 meV were used in the
test runs, and then the TZP basis set was chosen to perform the calculations
of cross sections, IMFP, and stopping power. 

In the LR-TDDFT 
calculations, $\textit{q}$ cannot take values smaller than the distance between two 
$k$-points in the BZ.
This distance can be estimated as $2\pi/(n_\mathrm{k} a)$, where $a$ is the 
lattice constant and $n_\mathrm{k}$ is the number of $k$-points in a 
particular direction. For a BZ sampling of $[3\times3\times3]$ 
$k$-points used in our calculations, we approximated the optical-limit by 
$\textit{q}_{0} = 0.1$ a.u. 
We have calculated ELF for a total of 20 values of $\textit{q}$ in the interval 
[0.1:2.0] a.u. The resolution in energy loss was defined by $\Delta E = 0.15$ eV and the broadening constant $\eta = 0.3$ eV (a sensible value must be $\geq2\Delta E$).

\subsection{\label{sec:water}Water samples}

\begin{figure}[h]
\centering
 \includegraphics[width=0.7\textwidth]{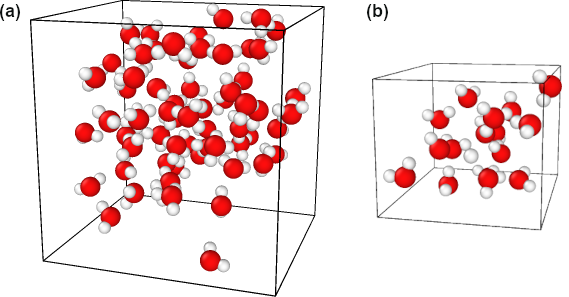}
 \caption{\label{fig:water} Unit cells of (a) PBE-64 and (b) PBE-16 water samples.}
\end{figure}

The results presented in sections~\ref{sec:ELF} and~\ref{sec:res-CS} were calculated for a water sample 
denoted as PBE-64 (referring to the exchange-correlation functional and the number of water 
molecules) shown in figure~\ref{fig:water}(a). The sample PBE-64 is composed of 64 water molecules which were initially randomly placed 
inside a cubic cell with a lattice constant $a = 12.45$ {\AA}, giving a density of 0.995 g/cm$^3$. The 
structure was equilibrated using Born-Oppenheimer molecular
dynamics (BOMD) using 
the {\sc siesta} code~\cite{siesta}. The BOMD simulations were performed
for a total time of 2.5 ps at a temperature of 300 K in the NVT ensemble (Nos{\'e} thermostat) with default Nos{\'e} mass of 100 Ry fs$^2$. The time step of 0.5 fs was used in the calculations. A double-$\zeta$ polarized (DZP) 
basis set of numerical atomic orbitals was used in the {\sc siesta} BOMD 
simulations~\cite{Artacho99,PhysRevB.64.235111}. The cut-off radii of 
the first-$\zeta$ functions were defined by an energy shift of 20 meV.
The second-$\zeta$ radii were defined by a split norm of 0.3. Soft 
confining potentials of 40 Ry with default inner radius of 0.9 were used in the basis-set generation 
\cite{PhysRevB.64.235111}. The plane-wave cutoff for the real-space grid 
was defined by a mesh cutoff of 300 Ry. The self-consistency was controlled
by a convergence parameter of $10^{-3}$ eV for the Hamiltonian matrix 
elements. The generalized gradient approximation (GGA) in the 
PBE~\cite{PhysRevLett.77.3865} form was used to account for the 
XC effects.

The calculations presented in sections~\ref{sec:analysis} and~\ref{sec:MO} 
were performed for the water sample PBE-16 (figure~\ref{fig:water}(b)) composed of 16 water molecules to reduce the computational cost. The sample
PBE-16 was optimized using BOMD with the same parameters as the sample PBE-64.
 
A few additional water samples of different sizes, atomic configurations, and equilibrated with different XC-functionals 
were used for testing purposes (see details in Appendix A).

\section{\label{sec:results} Results and discussion}

\subsection{\label{sec:ELF} Energy loss function}

\begin{figure}[h]
\centering
 \includegraphics[width=0.6\textwidth]{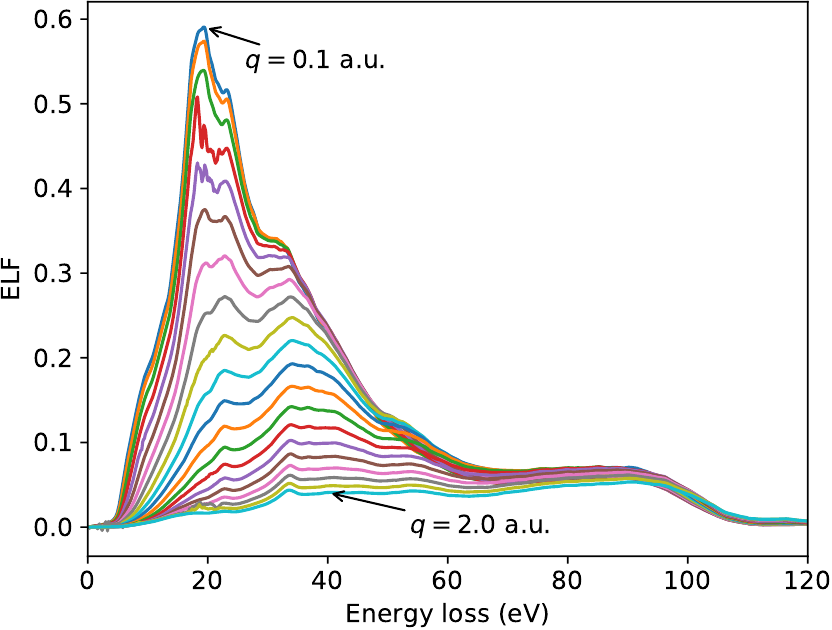}
 \caption{\label{fig:elf-all} ELF of liquid water (a.u.) as a function of energy loss (eV) calculated with LR-TDDFT for $\textit{q}=[0.1:2.0]$ a.u. PBE-64 water sample and TZP basis set were used in all calculations.}
 \end{figure}

The energy loss function for different values of the momentum transfer is 
shown in figure~\ref{fig:elf-all} as a function of energy loss. The 
ELF exhibits a clear evolution for different values of the momentum transfer.
At small $\textit{q}$, a defined feature (i.e., a single maximum, accompanied
by some shoulders) 
is clearly visible associated with the optical excitations sensitive to the 
optical band gap. 
For larger $\textit{q}$, the energy loss involves larger 
wave-vector excitations linked to the band structure of the system.

\begin{figure}[h]
\centering
 \includegraphics[width=0.99\textwidth]{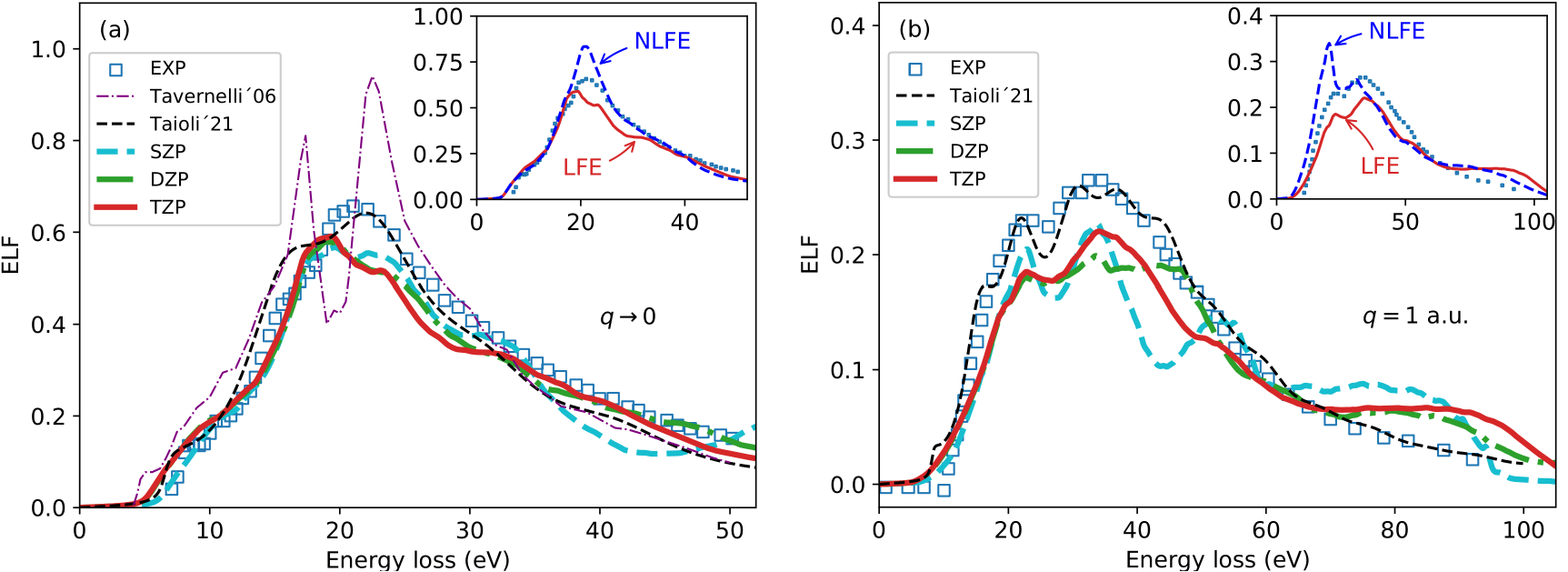}
 \caption{\label{fig:basis-q0} Energy loss function of liquid water: (a)
 in the optical limit $\textit{q} = 0.1$ a.u. and (b) for the momentum transfer
 $\textit{q} = 1.0$ a.u. as a function of energy loss (eV) calculated 
 with LR-TDDFT for PBE-64 water sample using different basis set sizes (SZP, DZP, and TZP) 
 in the {\sc siesta} calculations as indicated for each line. LR-TDDFT results
 are compared to the experimental data obtained via inelastic 
 X-ray scattering (IXS) spectroscopy data~\cite{haya2000} and to 
 calculations by Tavernelli \cite{PhysRevB.73.094204} (for $\textit{q} \rightarrow 0$) and Taioli et al.~\cite{Taioli2021}. Insets show the ELF with and without local field effects (LFE and NLFE, respectively) compared to experiment.}
 \end{figure}

Figure~\ref{fig:basis-q0} shows the comparison of our results for the 
momentum transfer of 0.1 and 1 a.u. with inelastic X-ray 
scattering (IXS) experimental data~\cite{haya2000}. A
convergence test for the ELF with respect to the basis set size is 
presented as well. The DZP and TZP results for the ELF at $\textit{q}=0.1$ a.u. are 
almost identical. At $\textit{q}=1.0$ a.u., the DZP and TZP slightly vary, with TZP 
closer reproducing main experimental features around the maximum of the ELF.
Main panels show the results obtained
including the LFE, while the insets compare the results obtained with
and without the LFE to the experimental data.

In the optical limit (figure~\ref{fig:basis-q0}(a)), the tails of the ELF 
from IXS experiments are well reproduced by our well-converged (TZP) calculations. The main peak is slightly underestimated and 
appears slightly shifted to lower energy loss values as
compared to the experiment. The results of Tavernelli 
\cite{PhysRevB.73.094204} obtained with real-time TDDFT (RT-TDDFT) are also 
shown in figure~\ref{fig:basis-q0}(a) and, as have been mentioned
before, the RT-TDDFT ELF shows two maxima with a minimum located at similar energy loss values as the experimental maximum.

Recent calculations by Taioli et al.~\cite{Taioli2021}
obtained with LR-TDDFT (including XC effects in the AGGA), are also shown in 
figure~\ref{fig:basis-q0}. The main peak is captured well by Taioli et 
al.~\cite{Taioli2021}, however, a feature at $E\approx15$ eV is higher 
than in the experimental ELF, and the main peak shows two features (although much less prominent) similar to the results of Tavernelli. Thus, our RPA results
better reproduce both the lower and higher energy-loss side of the main 
peak, while the AGGA results from Taioli et 
al.~\cite{Taioli2021} better reproduce the main peak.
Often, RPA with LFE is able to reproduce fairly well the energy loss 
spectra at $\textit{q}=0$~\cite{Botti_2004}.
AGGA generally brings an improvement upon RPA in finite systems~\cite{Gross2012}.
However, for extended non-metallic systems, the XC-kernel effect vanishes 
due to the absence of the long-range ($1/r$) decay~\cite{Gross2012,ghosez,kim} and thus AGGA yields a relatively small correction to the RPA results~\cite{PhysRevB.61.10149,PhysRevLett.86.5962}. The inclusion of the LFE, however, plays a significant role in the correct description of the electron energy loss~\cite{PhysRevB.61.10149}, as proven by our results.

At finite value of the momentum $\textit{q}=1$ a.u., positions of experimental
peaks and their widths are well captured by our calculations 
(figure~\ref{fig:basis-q0}(b)). However, the height is slightly underestimated 
and there is a plateau at high energy loss in the calculated ELF which is not 
present in the 
experimental data. Taioli et al.~\cite{Taioli2021} do not get such 
plateau in their calculations and they capture the height and width of 
the peaks quite accurately. Overall, this comparison clearly confirms, in
a TDDFT framework and considering the system beyond an electron gas as 
done in optical data models for TS calculations, that considering XC 
effects is more important for finite momentum transfer, as already 
anticipated by previous works in optical data models
~\cite{Emf2012}.

Apart from intensities and position of the peaks, one should also discuss
the "fall off" on the sides of the main structure. The extension 
algorithms used in many MC TS codes (e.g., in Geant4-DNA~\cite{inc2018}) do not account for the momentum broadening
of the ELF being largely based on the early Ritchie model(s) and the Ashley model, which 
exhibit a much steeper fall below the maximum. However, more recent MC TS codes have included momentum broadening either empirically~\cite{10.1667/RR13362.1} or via the Mermin dielectric function~\cite{Garcia_Molina_2011}.

In the insets to figure~\ref{fig:basis-q0}, we compare the ELF obtained with the local field effects (LFE), 
which is the same as the TZP results of the main panels, and without LFE 
(i.e., NLFE). 
In an inhomogeneous and polarizable system, on a microscopic 
scale, the LFE imply that the matrix $\epsilon_{\mathbf{GG}'}$ has non-zero 
off-diagonal elements. In practical terms, it means that the Coulomb 
interactions between the electrons of the system (i.e., the Coulomb kernel, Eq.~\eqref{kernel1}) are included.
The LFE are stronger when the inhomogeneity of the system is 
larger. Although, if the microscopic polarizability of the inhomogeneous 
system is small, the LFE are small.
For $\textit{q}\rightarrow 0$, the LFE only change the height of the main feature (inset to figure \ref{fig:basis-q0}(a)),
maintaining to a large extent the overall shape of the NLFE result. 
Practically, the LFE suppress the absorption, due to the induced 
classical depolarization potential. LFE are expected to give a small 
contribution in situations and systems with a smoothly varying electronic
density. On the contrary, the LFE give a more sizable contribution for larger wave-vectors as seen in the inset to figure~\ref{fig:basis-q0}(b).

\subsection{\label{sec:res-CS} Inelastic scattering cross sections and electronic stopping power}

Understanding the energy (and angle) distribution of secondary electrons 
is fundamental for characterizing the physical step of radiation 
bio-damage. Indeed, such energy will determine not only the specific 
processes by which such electrons will interact with the biological 
molecules in the water solvent, but also how close bond-breaking events, 
possibly occurring at nucleobase pairs or at the sugar-phosphate chain, 
occur (clustering). Depending on the clustering of these events, the 
damage may be irreparable. Here, we present the calculated 
single-differential as well as total inelastic scattering cross sections 
and compare our results with the calculations from the dielectric formalism. 
The IMFP and the stopping power are also compared to the results from the
default, Ioannina and CP100 models as implemented in Geant4-DNA.

\begin{figure}[h]
\centering
 \includegraphics[width=0.6\textwidth]{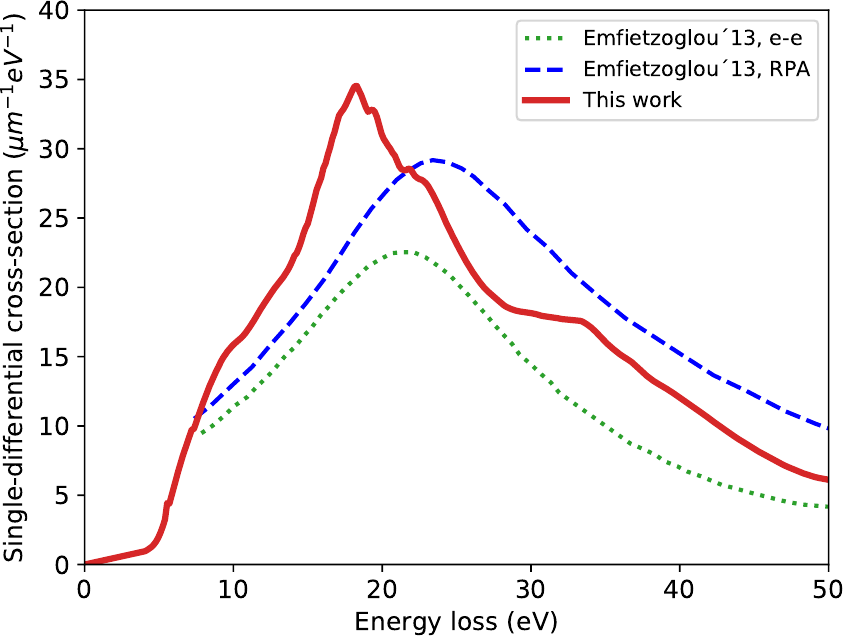}
  \caption{\label{fig:sdcs} Single-differential cross section 
 Eq.~\eqref{sdcs} for incident electron kinetic energy $T = 100$ eV. 
 Results from D. Emfietzoglou et al.~\cite{10.1667/RR13362.1} are shown 
 for comparison.}
 \end{figure}

Figure~\ref{fig:sdcs} shows the SDCS calculated using Eq.~\eqref{sdcs} for 
the electron kinetic energy $T=$ 100 eV in comparison
with different approximations of the Emfietzoglou model~\cite{10.1667/RR13362.1}. The label "e-e" in Emfietzoglou et 
al.~\cite{10.1667/RR13362.1} results, stands for the semi-empirical electron-electron 
dielectric function representing an exchange-correlation 
corrected screened interaction between the incident and struck electrons.  The label "RPA" 
stands for the random-phase approximation within the Lindhard formalism 
for dielectric function obtained
under the plasmon-pole approximation for a homogeneous electron gas. 
Further details can be found in Emfietzoglou et al.~\cite{10.1667/RR13362.1}. The 
maximum of SDCS obtained in this work is shifted to lower values of the 
energy loss as compared to the semi-empirical results. Overall, the 
agreement is qualitative. The SDCS for the electron kinetic energies of 500 eV, 1 keV, and 5 keV are given in Appendix A (figure \ref{fig:sdcs-difT}).

\begin{figure}[h]
\centering
  \includegraphics[width=0.6\textwidth]{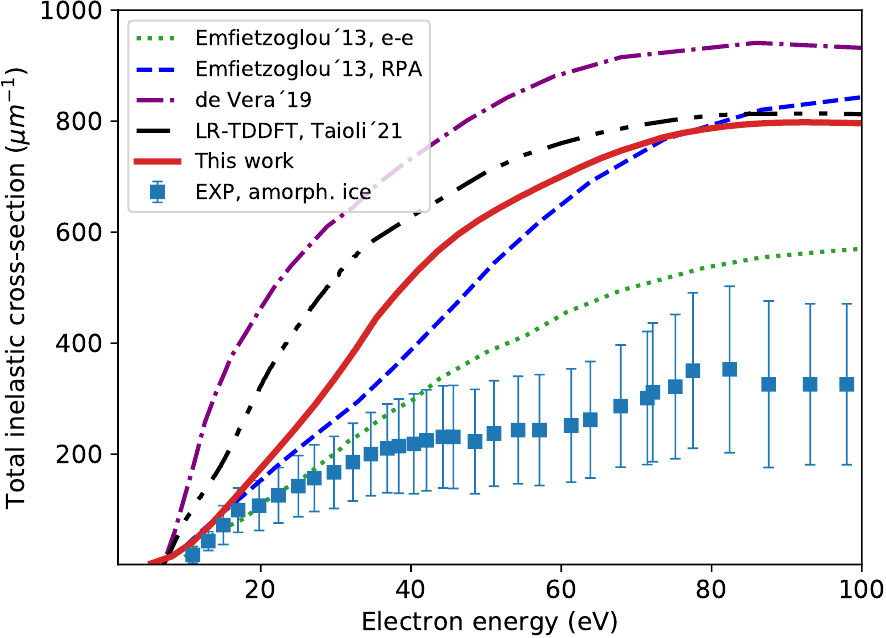}
\caption{\label{fig:tot-cs} Total inelastic scattering cross section 
Eq.~\eqref{tot_cs} as a function of the electron incident kinetic energy 
compared to experimental 
data for amorphous ice (EXP, amorph. ice)~\cite{10.1667/0033-7587(2003)159[0003:CSFLEE]2.0.CO;2} and 
calculations by Emfietzoglou et al.~\cite{10.1667/RR13362.1}, de Vera
et al.~\cite{deVera19}, and Taioli et al.~\cite{Taioli2021} (the sum of the excitation and ionization cross sections presented in figure 3 of Ref.~\cite{Taioli2021}).}
\end{figure}

The total inelastic cross section is shown in figure~\ref{fig:tot-cs} and compared to the results of Emfietzoglou et 
al.~\cite{10.1667/RR13362.1}, de Vera et al.~\cite{deVera19}, and Taioli et al.~\cite{Taioli2021}, as well 
as with experimental data for amorphous ice. Here, the total cross section is expressed in the units of inverse length, i.e., the conventional units of length squared multiplied by the number density of the target atoms~\cite{10.1667/RR13362.1}. Our result agrees with the RPA model 
of Emfietzoglou et al. being slightly higher at intermediate electron 
energies. Both RPA calculations, our LR-TDDFT and the Emfietzoglou model, converge to the experimental curve only at energies below 20 eV. Recent LR-TDDFT results from Taioli et al.~\cite{Taioli2021} are slightly higher than ours, but become similar at energies above 80 eV.
 
\begin{figure}[h]
 \centering
  \includegraphics[width=0.6\textwidth]{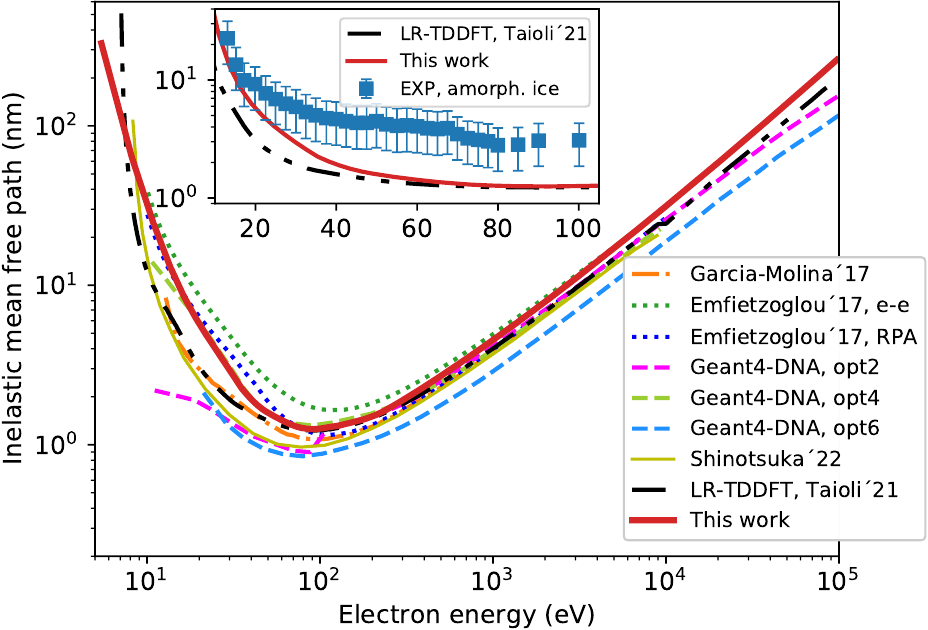}
   \caption{\label{fig:imfp} Inelastic mean free path as a function of the 
 electron incident kinetic energy compared to calculations by 
 Garcia-Molina et al.~\cite{garcia2017_49} (dash-dotted line), 
 Emfietzoglou et al.~\cite{https://doi.org/10.1002/sia.5878} (dotted 
 lines), Geant4-DNA options 2, 4, and 6~\cite{inc2018} 
 (dashed lines), Shinotsuka et al.~\cite{Shinotsuka2022}, and Taioli et al.~\cite{Taioli2021} (dash-dot-dotted line, calculated as the inverse of the total cross section shown in figure~\ref{fig:tot-cs}). The inset shows the comparison of the LR-TDDFT results from this work and from Taioli et al.~\cite{Taioli2021} with
 experimental data for amorphous ice~\cite{10.1667/0033-7587(2003)159[0003:CSFLEE]2.0.CO;2}.}
 \end{figure}

The IMFP obtained in this work is shown in figure~\ref{fig:imfp} and 
compared to the semi-empirical 
calculations~\cite{garcia2017_49,https://doi.org/10.1002/sia.5878}, the 
LR-TDDFT results from Taioli et al.~\cite{Taioli2021}, and to three 
Geant4-DNA constructors (the default, the Ioannina, and the CPA100 models, 
denoted as opt2, opt4 and opt6)~\cite{inc2018}. A more 
recent semi-empirical result of Shinotsuka et al.~\cite{Shinotsuka2022} 
obtained using the relativistic full Penn algorithm that includes the correction of 
the bandgap effect in water is also shown in figure~\ref{fig:imfp}. Our 
results quantitatively agree with the RPA model of Emfietzoglou et 
al.~\cite{https://doi.org/10.1002/sia.5878} in the whole energy range and 
with the IMFP obtained by Taioli et al.~\cite{Taioli2021} at intermediate 
energies (from 100 eV to 10 keV). The data of Shinotsuka et al.~\cite{Shinotsuka2022} is below our IMFP and resembles more the shape of the results of Taioli et al.~\cite{Taioli2021}, except for the region around the minimum. The inset of figure~\ref{fig:imfp} shows 
the comparison of the LR-TDDFT results from this work
and from Taioli et al.~\cite{Taioli2021} with the experimental data for 
amorphous ice below 100 eV. Both calculated results agree well at energies 
above 50 eV. Below this energy, the results of this work are closer to the 
experimental data than the ones from Taioli et al.~\cite{Taioli2021}.

In the Geant4-DNA default option (opt2), the 
total and inelastic cross sections for weakly-bound electrons are 
calculated from the energy- and momentum-dependent complex dielectric 
function within the first Born approximation. In particular, the optical 
data model of Emfietzoglou et al.~\cite{emfi2355,emfi2005} is used, where
the frequency dependence of the dielectric function at $\textit{q}=0$ is obtained 
by fitting experiments for both the real and imaginary part of the 
dielectric function, using a superposition of Drude-type functions with 
adjustable coefficients.
A partitioning of the ELF to the electronic absorption channels at $\textit{q}=0$,
proportional to the optical oscillator strength, enables the calculation 
of the cross sections for each individual excitation and ionization
channel. The extension to the whole Bethe surface is made by 
semi-empirical dispersion relations for the Drude coefficients. Below a 
few hundred eV, where the first Born approximation is not applicable, a 
kinematic Coulomb-field correction and Mott-like XC 
terms are used~\cite{emfi2005}. For ionization of the O K-shell, total 
and differential cross sections are calculated analytically using the 
binary-encounter-approximation with exchange model (BEAX), an atomic 
model which depends only on the mean kinetic energy, the binding energy, 
the occupation number of the electrons, and where the deflection angle is
determined from the kinematics of the binary collision, thus referring to
sole vapour data.

In the Ioannina model~\cite{ioannina,kyr2015}, two problems appear in the default
option, e.g., a brute-force truncation of the Drude function violating 
the $f$-sum rule and the consequent complexity in deriving 
$\epsilon_R(E,\textit{q})$ from $\epsilon_I(E,\textit{q})$ via Kramers-Kronig 
relations~\cite{inc2018} are overcome via an algorithm which 
redistributes $\epsilon_I(E,\textit{q}=0)$ to the individual inelastic channels
in a $f$-sum rule constrained and physically motivated manner. Below a 
few hundred eV, more accurate ionization cross sections, especially at 
energies near the binding energy, are obtained via methodological 
improvements of the Coulomb and Mott corrections. In the CPA100 
models~\cite{bordage}, excitation cross sections are calculated in 
the first Born approximation using the optical data model by Dingfelder 
et al.~\cite{ding53}, which is also based on a Drude-function 
representation of $\epsilon(E,\textit{q})$ but uses a different 
parametrization. The ionization cross sections are calculated, via the Binary-encounter-Bethe (BEB) atomic model.

As no international recommendations exist yet for the mean free path, the
only conclusion we can draw from the comparison between our results and 
the Geant4-DNA (figure~\ref{fig:imfp}) is that our first-principles result seems to well 
reproduce the Geant4-DNA opt4 at energies up to 1 keV. 
Above this energy, our result is slightly higher than opt2, opt4 and 
opt6. For the whole energy range,
the IMFP from opt6 appears to be the lowest because of the larger 
inelastic cross sections in the 10 eV-10 keV range, as a consequence of using an atomic ionization model with the absence of screening~\cite{inc2018}. The 
curve for opt2 shifts suddenly, through a clear visible step, below 200 eV because of the activation of vibrational 
excitations, which reduce additional energy losses and thus reduce the 
IMFP.
In opt4, excitations are strongly enhanced compared to ionization, 
the latter decreasing only moderately, which results in higher $W$ values
(the average energy to produce an ion pair) and smaller penetration
distances~\cite{ioannina}.
Since XC effects mostly affect the results for $\textit{q}\neq$0, it is expected 
that XC corrections to the RPA will mostly influence the IMFP at low 
energies, where large-angle scattering collisions ($\textit{q}\neq$0) become
important~\cite{https://doi.org/10.1002/sia.5878,10.1667/RR13362.1}.
Indeed, as the comparison in the inset of 
figure~\ref{fig:imfp} shows, the RPA result of this work differs from 
the AGGA results of Taioli et al.~\cite{Taioli2021} only at energies below 
50 eV. However, the inclusion of exchange and correlation does not improve 
the RPA result with respect to the experimental data.

\begin{figure}[h]
 \centering
   \includegraphics[width=0.6\textwidth]{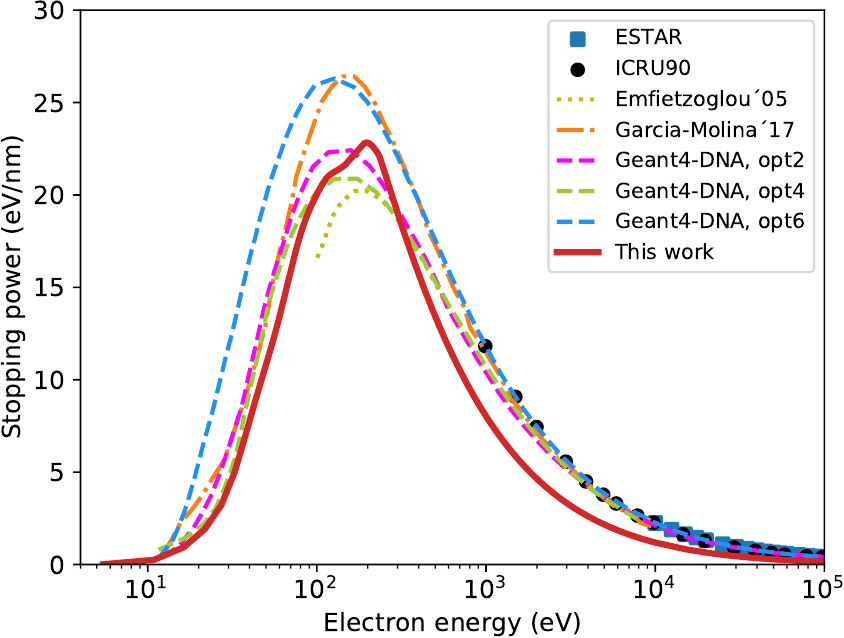}
 \caption{\label{fig:stop} Electronic stopping power Eq.~\eqref{stopp}
 as a function of the electron kinetic energy. Our results are compared 
 to ESTAR~\cite{estar} and ICRU90~\cite{icru90} data, as well as 
 calculations by Emfietzoglou et al.~\cite{emfi2005bis} (dotted line), Garcia-Molina et al.~\cite{garcia2017_49} (dash-dotted
 line), and Geant4-DNA options 2, 4, and 6~\cite{inc2018} 
 (dashed lines).}
 \end{figure}

The stopping power from LR-TDDFT is shown in figure~\ref{fig:stop} in 
comparison with the semiempirical 
calculation~\cite{emfi2005bis,garcia2017_49}, three Geant4-DNA options, 
and data from ICRU and ESTAR. As it
was the case for IMFP, our result is closer to opt4, until a few hundreds
of eV, while at higher energies
our stopping power is considerably smaller than the rest of the data 
presented. However, our stopping recovers the correct limit at the 
highest energies (10-100 MeV).

\subsection{\label{sec:analysis} Analysis of different contributions to the electron energy loss function}

 Following the partition method described in Sec.~\ref{sec:partition}, we 
calculated the ELF separated on contributions from different species, 
angular momenta, and atomic pairs.

\begin{figure}
 \centering
 \includegraphics[width=0.98\textwidth]{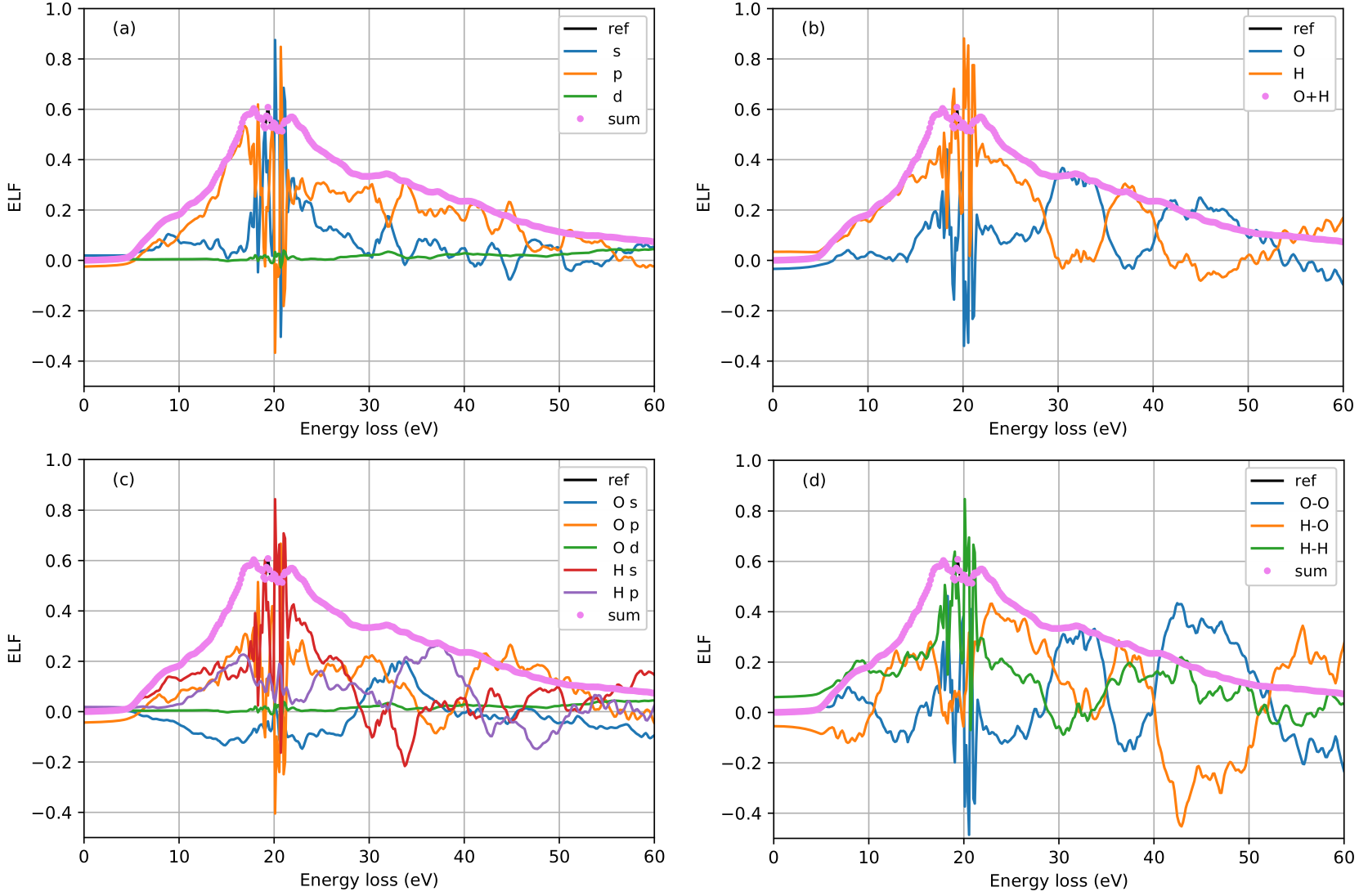}
 \caption{\label{fig:ELF-contrib} Contributions of (a) different angular momenta; (b) species; (c) angular momenta of different species; and (d) pairs of species, to the total ELF ($\textit{q}=0.1$ a.u.) labeled as "ref". }
\end{figure}

Figure~\ref{fig:ELF-contrib}(a), shows the contribution of orbital angular 
momenta $s$, $p$, and $d$ of all the atoms to the total ELF. Clearly, 
$p-$orbitals play a major role in forming the slopes of the ELF with a smaller 
contribution from $s-$orbitals. The maximum of the total ELF comes from a
complicated interplay between the $s-$ and $p-$levels, for which the ELF 
oscillates in an incoherent way. The results are validated by summing up 
all the contributions which add up to the total ELF.

Figure~\ref{fig:ELF-contrib}(b) shows the contribution of all the oxygen atoms 
and all the hydrogen atoms to the total ELF. Below the maximum, the 
total ELF mostly comes from the hydrogen atoms. At the maximum, hydrogen
atoms contribute more to the total ELF, while both curves again oscillate
in opposite phases. Above the maximum, the two contributions oscillate
resulting in a rather smooth total slope.

A detailed analysis, separating $s$, $p$, and $d$ orbitals of hydrogen 
and oxygen (figure~\ref{fig:ELF-contrib}(c)), shows that the main contribution
to the ELF maximum is from the hydrogen $s-$levels and oxygen $p-$levels.
Oxygen $s$ orbitals mostly contribute negatively. A significant
contribution comes from the hydrogen $p$ orbital, indicating the 
excitation of the hydrogen electrons which populate the $p-$shell. Oxygen
$d-$shell remains mostly unpopulated.
%

The contributions of species pairs are shown in figure~\ref{fig:ELF-contrib}(d).
Again, hydrogen plays the main role at low energy loss as H-H and H-O 
pairs. At the maximum of ELF, both H-O and O-O pairs contribute
negatively to the total ELF. Similarly to the case of different species, 
there are incoherent oscillations in ELF above 25 eV for O-O and H-O 
pairs. However, in this case, the two contributions oscillate around zero
cancelling each other. Thus, at high energy loss, the total ELF is mostly
due to the H-H pairs.

\subsection{\label{sec:MO} ELF and SDCS for each molecular orbital}

When constructing the Drude-type dielectric response function in 
semi-empirical methods, the continuum in the fitting procedure is usually 
represented by the outer shells of the water molecule 
\cite{EMFIETZOGLOU200271,EMFIETZOGLOU2003373}. Thus, the analysis of the ELF, and consequently 
the cross-sections, for each molecular orbital of water
can be of interest for the MC track structure community for benchmarking
of the semi-empirical models.

\begin{figure}[h]
\centering
 \includegraphics[width=0.7\textwidth]{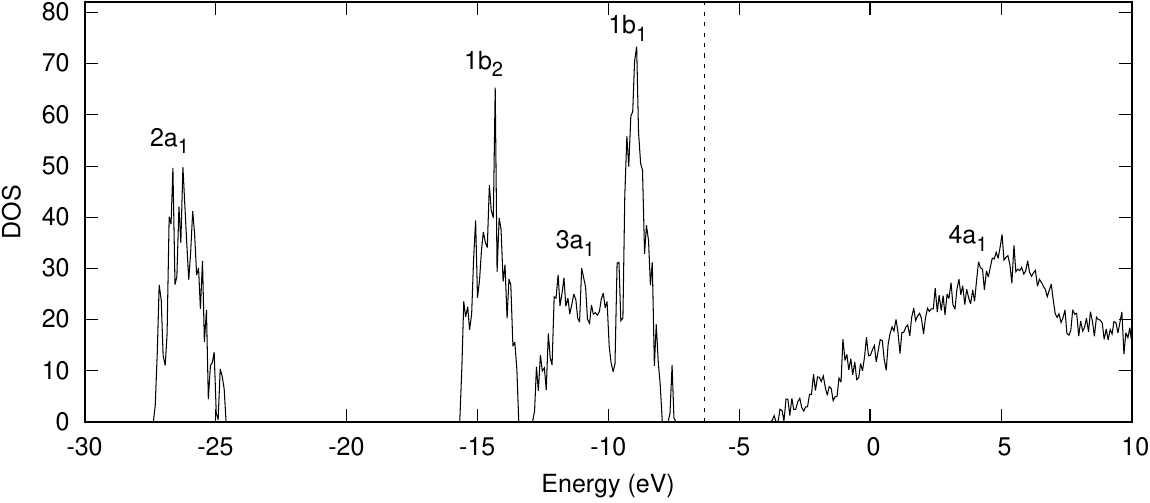}
 \caption{\label{fig:dos} Electronic density of states of liquid water obtained within DFT. Vertical dashed line shows the Fermi level at $-6.3$ eV.}
\end{figure}

Here, we calculate the ELF for each occupied molecular orbital of the water 
molecule, i.e., the orbitals 2a$_1$, 1b$_2$, 3a$_1$, and 
1b$_1$~\cite{doi:10.1021/acs.chemrev.7b00259}.
In liquid water, the bands of crystal orbitals 
correspond to different symmetries of an isolated molecule. The bands can
be seen in the electronic density of states (DOS) of liquid
water sample shown in figure~\ref{fig:dos} as a function of energy which
we calculated using the DFT implementation
of the {\sc siesta} code~\cite{siesta}. The binding energies of the four 
occupied orbitals of water are known from photoemission experiments.
For liquid water, the binding energies are 30.90 eV for 
2a$_1$, 17.34 eV for 1b$_2$, 13.50 eV for 3a$_1$, and 
11.16 eV for 1b$_1$~\cite{doi:10.1021/jp030263q}. Our 
results are slightly higher than experimental data, 
which is expected from the DFT method. However, 
our DOS is in good agreement with other DFT calculations~\cite{doi:10.1063/1.1940612}.

Each feature in the DOS (figure~\ref{fig:dos}) is labeled with the symmetry group corresponding to the isolated molecule. The DOS includes four outer occupied
bands with symmetries 2a$_1$, 1b$_2$, 3a$_1$, and 1b$_1$ and one unoccupied 
band with the symmetry 4a$_1$. For the ELF calculations, we only considered
the occupied orbitals, i.e., the ones located below the Fermi level.

\begin{figure}[h]
\centering
 \includegraphics[width=0.6\textwidth]{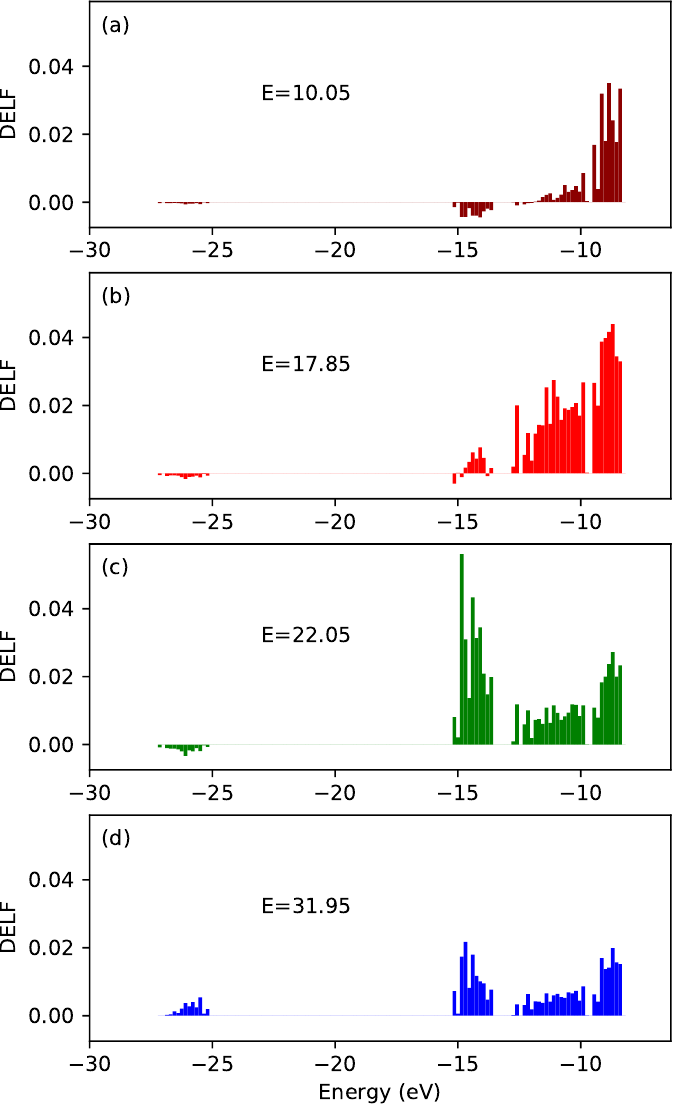}
 \caption{\label{fig:ELF-occ-en} Contributions of different occupied states 
 to the total ELF ($\textit{q}=0.1$) resolved in energy
 shown on panels (a), (b), (c), and (d) for the energy loss $E$ values 10.05, 17.85, 22.05, and 31.95 eV, respectively.
}
\end{figure}

Since we cannot directly obtain the ELF for each molecular orbital from 
LR-TDDFT calculations, we sum up the values of ELF for all occupied crystal
orbitals within the energy window corresponding to each symmetry.
Figure~\ref{fig:ELF-occ-en} shows the contributions of 
occupied states to the total ELF resolved in energy
in the optical limit for four selected values of the energy loss 
$E=10.05$, 17.85, 22.05, and 31.95 eV, corresponding to the regions below, 
around, and above the maximum of ELF (see figure~\ref{fig:basis-q0}(a)).
One can clearly distinguish four energy windows in which the ELF has 
non-zero values that can be directly correlated with the DOS 
(figure~\ref{fig:dos}). Thus, summing up the ELF of each crystal orbital in 
each of the energy windows, we obtained the ELF for four molecular orbitals 
of water. 

\begin{figure}[h]
\centering
 \includegraphics[width=0.98\textwidth]{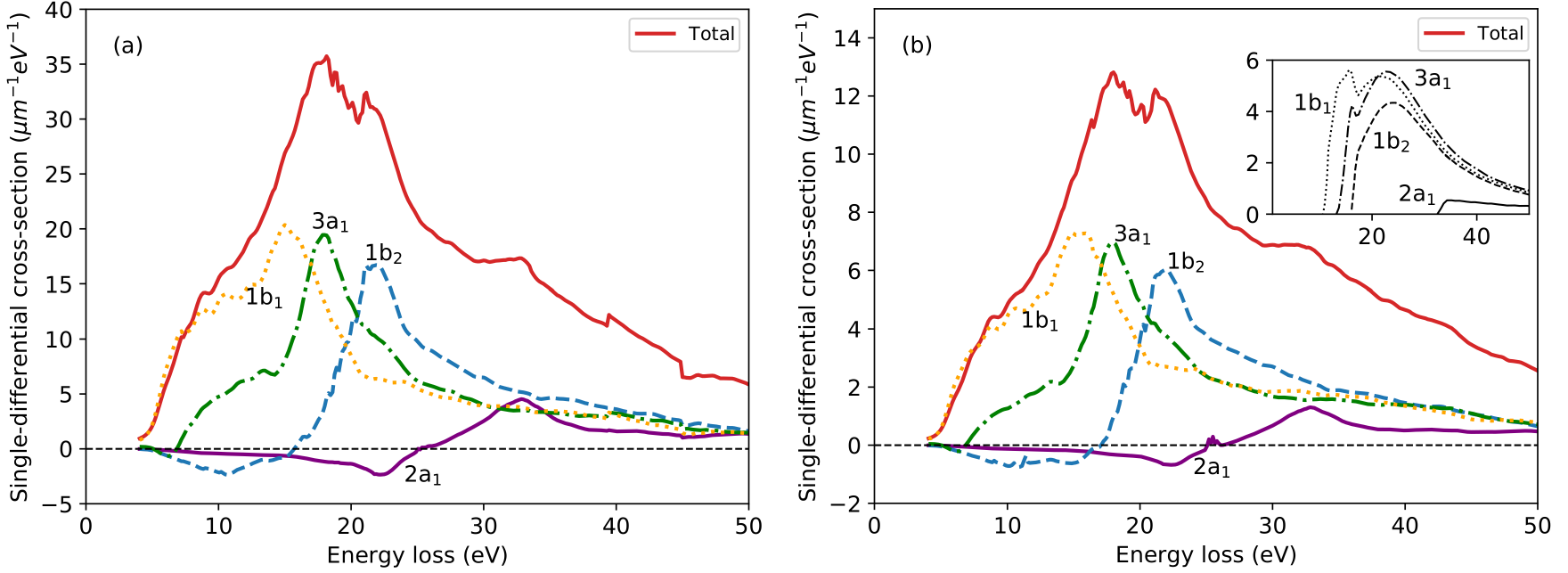}
 \caption{\label{fig:sdcs-orb} Single-differential cross section 
 for each symmetry of a water molecule. Incident electron kinetic energy is (a) $T = 100$ eV and (b)  $T = 500$ eV. The inset in panel (b) shows the results from Ref.~\cite{ding53}.}
 \end{figure}

We repeated the calculations described above for the same 
values of the momentum transfer $q$ as in figure~\ref{fig:elf-all} to obtain
the partitioned ELF in the whole Bethe surface ($E,\textit{q}$). This allowed us to 
compute the cross sections corresponding to each molecular orbital 
using Eq.~\eqref{sdcs}. 
As an example, figure~\ref{fig:sdcs-orb} shows the single-differential cross
sections for the incident electron with kinetic energies $T=$
100 eV and 500 eV in water sample with 16 molecules. Each molecular orbital
is observed to be responsible for a certain feature in the total SDCS.
The partitioning also looks very similar at both electron incident energies.
The cross section for energy losses below 10 eV is almost entirely due
to the contribution from the highest occupied molecular orbital (HOMO) 
1b$_1$. Deeper shells contribute at higher energy losses. The orbital 2a$_1$
has a minor contribution only at high energy 
losses. The observed behavior is in a qualitative agreement with available 
data from the dielectric formalism within the first Born 
approximation~\cite{ding53,EMFIETZOGLOU2003373} (see inset to figure~\ref{fig:sdcs-orb}(b)).

\section{Conclusions}
 
In summary, in this work, we computed several quantities important for the 
description of inelastic scattering of electrons in liquid water using a 
linear-response formulation of TDDFT. A good agreement with experimental data
was obtained for ELF in the optical limit as well as at finite values of the 
momentum transfer. We thoroughly tested our results
for dependence on the system size and the choice of the DFT parameters. 

Additionally, we provided a detailed analysis of the ELF in the optical limit
in terms of contributions from different species, species pairs, and orbital 
angular momenta.

Furthermore, we computed the single-differential cross section, total
inelastic cross section, inelastic mean free path, and the electronic 
stopping power from the ELF. Our results are in a good agreement
with the semi-empirical calculations. Thus, LR-TDDFT offers an 
alternative method to the standard semi-empirical calculations and
provides useful input for more detailed Monte Carlo track structure 
simulations. It is envisioned that the investigated quantities have the 
potential to be of direct use in open source TS codes like Geant4-DNA.
In particular, the decomposition of the cross sections on different 
molecular orbital channels, calculated \emph{ab initio} for the first time in
this work can be used as a benchmarking for semi-empirical models.

\section*{Appendix A. Test results and additional data for SDCS}

To check the convergence of our results with respect to the particular 
configuration of the water molecules in the sample, we have calculated the
ELF for several different samples. One of the samples with the initial 
configuration of PBE-64 was equilibrated with the Van der Waals 
(DRSLL)~\cite{PhysRevLett.92.246401,PhysRevLett.103.096102} 
XC-functional in the BOMD simulation. Three snapshots were chosen, after 
2.7 ps, 5.0 ps, and 6.8 ps of the BOMD, denoted with subscripts 
VdW-64$\_${$1,2,3$} to test the convergence of the LR-TDDFT results with the
sample configuration.
The sample RPBE-64 (RPBE-128) contains 64 (128) molecules in a cubic 
unit cell with the lattice constant $a = 12.43$ $(15.66)$ {\AA} to 
correspond to the experimental water density at 300 K and 1 bar. The SCF 
convergence threshold during the MD was st to $2\times10^-7$. The 
equilibration was performed with BOMD of the CP2K code~\cite{doi:10.1063/5.0007045}, with 
RPBE-D3 XC-functional and a TZV2P basis set at 300 K in the NVT ensemble 
(Nos{\'e}-Hoover-chains thermostat) for 10 ps. A
larger sample with 128 molecules was used to test the dependence of our 
results on the system size. 

\begin{figure}[h]
\centering
 \includegraphics[width=0.6\textwidth]{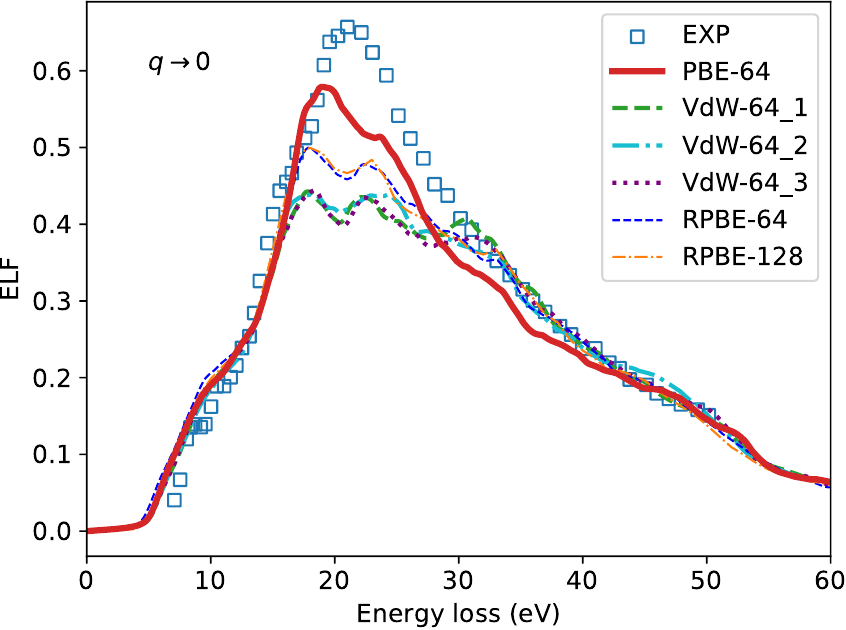}%
 \caption{\label{fig:sizes} Energy loss function of liquid water in the 
 optical limit ($q\rightarrow0$) as a function of energy loss (eV) 
 calculated with LR-TDDFT for different water structures (see text for 
 details) with the broadening constant $\eta=0.3$ eV and the momentum 
 transfer $q=0.1$ a.u. DZP basis set with an energy shift of 20 meV is 
 used in the {\sc siesta} calculations. Inelastic X-ray scattering (IXS) data 
 ($q = 0$) is from Hayashi et al.~(Ref~[31] of the main text).}%
\end{figure}

The tests have shown that the ELF is similar for
all the structures and that it does not depend on the system size or a 
particular snapshot configuration among the ones generated as described, 
once properly equilibrated (figure \ref{fig:sizes}). The three different VdW
snapshots give practically the same results. The same 
is true for both RPBE samples, with 64 and 128 molecules, the use of which leads to the same results for the ELF. 
Overall, the tests show that the ELF does not
depend on neither the size of the sample used in the calculations, nor
the exact configuration of the water molecules in the sample. The 
choice of the XC functional, however, affects the ELF results in the peak area, with the PBE functional giving the closest result to the experimental data. For this reason,
we chose to perform all the calculations for the cross sections and 
stopping power using the ELF obtained with the PBE-64 structure.

\begin{figure}[h]
\centering
 \ \ \ \ \ \includegraphics[width=0.6\textwidth]{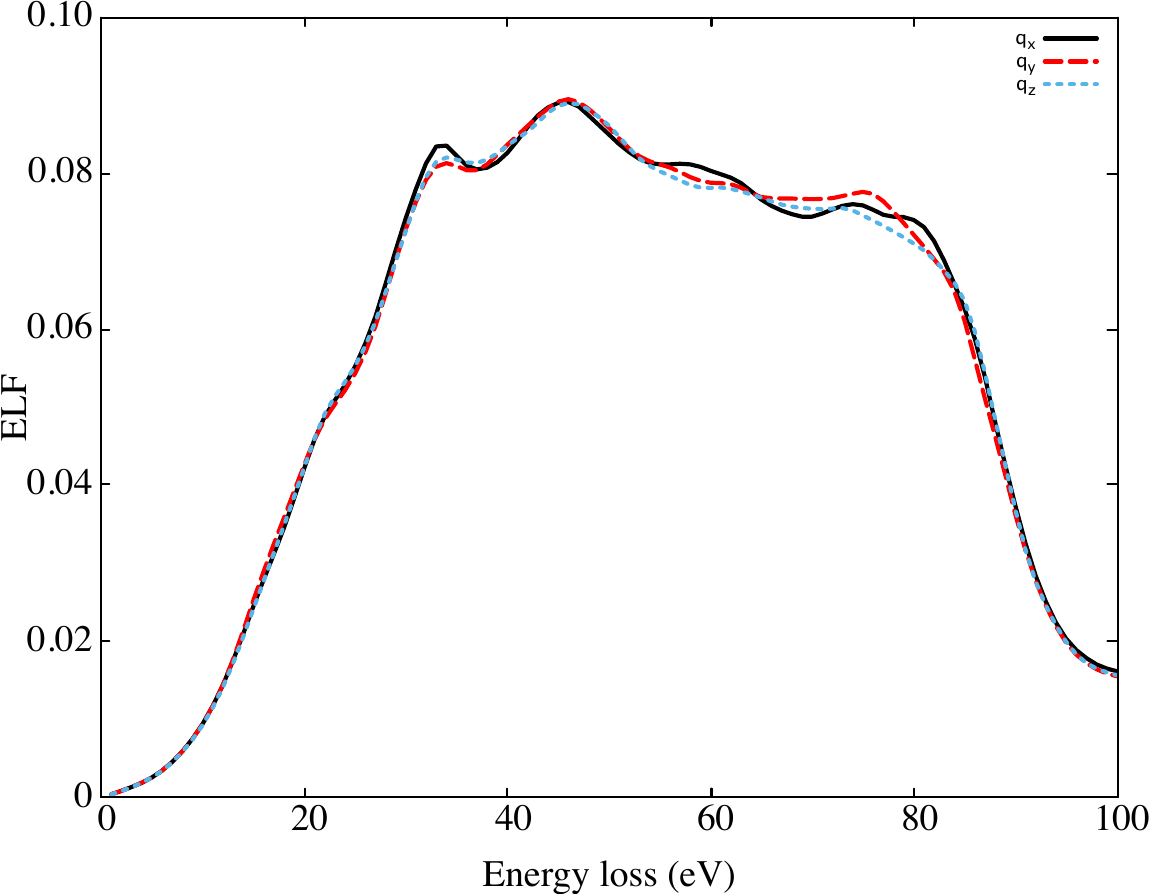}%
 \caption{\label{qdir} Energy loss function of liquid water (sample PBE-64) for $q=1.5$ a.u. as a function of
 energy loss (eV) 
 calculated with LR-TDDFT for different directions of the momentum 
 transfer: $q_x$, $q_y$, and $q_z$.}%
\end{figure}

The use of the scalar value of the momentum transfer is justified by the fact that the ELF is isotropic, i.e., it is very similar for all the directions of the momentum transfer vector, as can be 
seen in figure~\ref{qdir}.

\begin{figure}[h]
\centering
\includegraphics[width=0.57\textwidth]{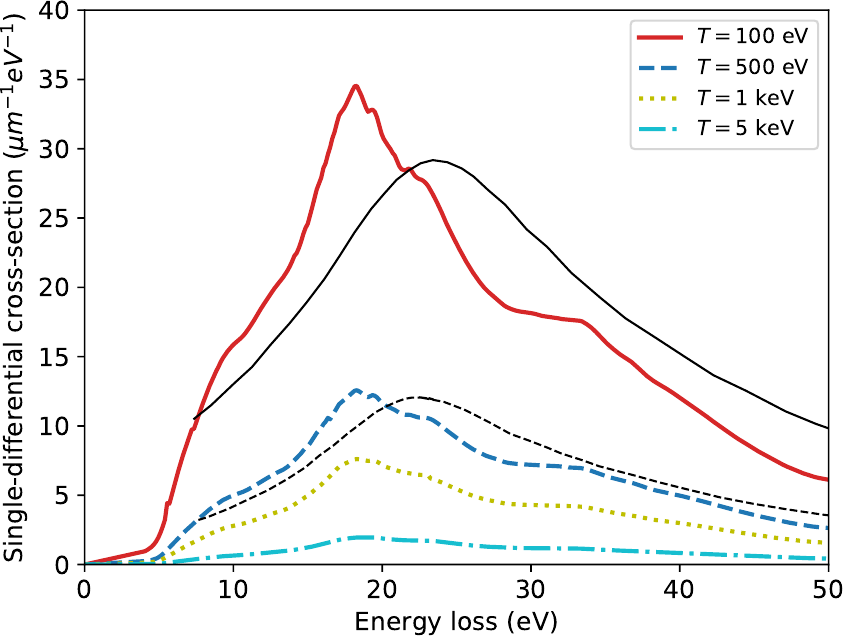}
\caption{\label{fig:sdcs-difT} The SDCS for the electron kinetic energies of 100 eV, 500 eV, 1 keV, and 5 keV as a function of energy loss. Thin black lines show the RPA results from Emfietzoglou et al. \cite{10.1667/RR13362.1} for electron energies of 100 eV (solid black line) and 500 eV (dashed black line).}
\end{figure}

Figure \ref{fig:sdcs-difT} presents the single-differential cross sections
for incident electron kinetic energies of 100 eV, 500 eV, 1 keV, and 5 keV compared to available results from Emfietzoglou et al. \cite{10.1667/RR13362.1} for energies of 100 and 500 eV.

\begin{acknowledgments}
The presented work has been funded by the Research Executive Agency under 
the European Union's Horizon 2020 Research and Innovation program (project
ESC2RAD: Enabling Smart Computations to study space RADiation effects,  Grant  Agreement 776410). JK was supported by the Beatriz 
Galindo Program (BEAGAL18/00130) from the Ministerio de Educaci\'on y 
Formaci\'on Profesional of Spain, and by the Comunidad de Madrid through 
the Convenio Plurianual with Universidad Polit\'ecnica de Madrid in its 
line of action Apoyo a la realizaci\'on de proyectos de I+D para 
investigadores Beatriz Galindo, within the framework of V PRICIT (V Plan 
Regional de Investigaci\'on Cient\'ifica e Innovaci\'on Tecnol\'ogica). EA
acknowledges the funding from Spanish MINECO through
grant FIS2015-64886-C5-1-P, and from Spanish MICIN
through grant PID2019-107338RB- C61/AEI/10.13039/501100011033, 
as well as a Mar\'{\i}a de Maeztu award to Nanogune, Grant 
CEX2020-001038-M funded by MCIN/AEI/ 10.13039/501100011033. We thank Simone Taioli and Pablo de Vera for providing us with the data related to their recent publications. We are grateful for computational resources provided by 
Donostia International Physics Center (DIPC) Computer Center and 
Barcelona Supercomputer Center. We thank Dr. Daniel Mu\~noz Santiburcio 
for providing the water structures obtained by CP2K calculations.
\end{acknowledgments}


%
%

%

\clearpage
\bibliography{biblio-1.bib}

\end{document}